\documentstyle[aps,floats,twocolumn,prl]{revtex}
\def\beq{\begin{equation}}
\def\eeq{\end{equation}}
\def\beqa{\begin{eqnarray}}
\def\eeqa{\end{eqnarray}}
\newcommand{\beqas}{\begin{eqnarray*}}
\newcommand{\eeqas}{\end{eqnarray*}}
\def\sign{\mbox{sign}}
\def\oneh{\frac{1}{2}}

\def\wec{{\vec{w}}}

\def\<{\langle}
\def\>{\rangle}
\def \R{{\rm I\kern -0.13em R}}
\def \N{{\rm I\kern -0.13em N}}
\def \D{{\rm I\kern -0.13em D}}
\newcommand{\vek}[1]{\mbox{\boldmath$#1$}}
\newcommand{\smvek}[1]{\mbox{\boldmath${\scriptstyle #1}$}}

\newcommand{\ord}[1]{{\cal O}\left(#1\right)}

\newcommand{\mmin}[1]{\parbox[t]{2em}{\scriptsize{\normalsize \rm min} \\ 
	\centerline{$#1$}}}
\newcommand{\mmax}[1]{\parbox[t]{2em}{\scriptsize{\normalsize \rm max} \\ 
	\centerline{$#1$}}}

\begin{document}

\tighten\draft

\twocolumn[\hsize\textwidth\columnwidth\hsize\csname@twocolumnfalse\endcsname

\title{Statistical Mechanics of Support Vector Networks}
\author{Rainer Dietrich$^{1}$, Manfred Opper$^{2}$ and Haim Sompolinsky$^{3}$}
\address{
$^{1}$ Institut f{\"u}r Theoretische Physik,
Julius-Maximilians-Universit{\"a}t, Am Hubland, D-97074 W{\"u}rzburg,  
Germany.
\\
$^{2}$ Department of Computer Science and Applied Mathematics, Aston
University, Birmingham B4 7ET, UK.
\\
$^{3}$ Racah Institute of Physics and Center for Neural Computation,  
Hebrew University, Jerusalem 91904, Israel.
}

\maketitle

\begin{abstract}
Using methods of Statistical Physics, we investigate the
generalization performance of support vector machines (SVMs),
which have been recently introduced as a general alternative
to neural networks. For nonlinear classification rules,
the generalization error saturates on a plateau, when the number of
examples is too small to properly estimate the coefficients
of the nonlinear part. When trained on simple rules,
we find that SVMs overfit only weakly. The performance of SVMs
is strongly enhanced, when the distribution of the inputs has a gap
in feature space.
\end{abstract}

\pacs{PACS numbers: 87.10.+e.,05.90.+m} ]

\narrowtext
Statistical Mechanics provides an important approach
to analyzing and understanding the ability of neural networks
to learn and generalize from examples (see e.g.
\cite{Seu91,Waal93,OpKi}). The majority of this
work has been devoted to the simplest network
architecture, the perceptron. This network however has
limited power, because it classifies examples with a simple
linear separating hyperplane and is able to learn
only linear separable rules. More complicated multilayer neural
nets can realize general nonlinear rules (when the size
of their hidden layer is large enough) but have also practical
and theoretical disadvantages. Learning in these networks
results in a usually nonconvex optimization problem and there
is no guarantee that an algorithm will find the minimum
of the training cost function. The complexity of the training error
surface reflects itself in the theoretical analysis by Statistical
Mechanics. The occurence of phases of broken ergodicity
\cite{Monass} makes their analysis a complicated task.
Finally, network parameters must be chosen carefully in order to
adapt the network's complexity on the task and to avoid overfitting.

Recently, a new type of learning machine has been introduced
by V. Vapnik and his collaborators \cite{Vap,Cort} which may become a
reasonable alternative to neural networks. These
{\em support vector machines} (SVMs)
seem to have several advantages over neural networks.
Being generalizations of perceptrons, their training
involves only simple convex optimization. Further,
for several applications, it has been shown
that SVMs do not have a strong tendency to overfit.

In this letter, we present a detailed analysis of the typical
performance of SVMs by methods of Statistical Mechanics.
To understand the basic idea behind the SVM approach, assume
a nonlinear mapping $\vec{\Psi}(\vek{x})$ from
vectors $\vek{x}\in \R^N$ onto vectors $\vec{\Psi}$ which belong
to an $M$-dimensional feature space. A
nonlinear classification of inputs $\vek{x}$ can be defined
by a linear separation of feature vectors $\vec{\Psi}(\vek{x})$
using a perceptron with weight vector $\wec\in \R^M$
perpendicular to the separating hyperplane via
$\sign \left( \vec{\Psi}\left(\vek{x}\right) \cdot \vec{w} \right)$.
The dot denotes the standard inner product of vectors in $\R^M$.
The vector $\wec$ can be adapted to a set of example data by any
learning algorithm for perceptrons. This simple approach has
major problems which result from the typical
high dimensionality of the feature space. Assuming e.g.,
that the vector $\vec{\Psi}$
contains all bilinear expressions of components of the input vector
$\vek{x}$
(in addition to linear ones), the dimension $M$ is of
order $N^2$. First, there is a
big computational problem in storing and learning the weights and
second, one can expect that there is also a large tendency of these
machines to overfit, because there are much less training data
than adjustable parameters in this model.
The main idea to overcome these problems is to use
the optimal stability learning algorithm, which
has also been studied extensively in the Statistical Mechanics approach to
neural networks (see e.g. \cite{OpKi}).
The goal of this algorithm is to find a
vector of weights $\wec$
which allows for a separation of
positive and negative example points with
the maximal margin defined by
\beq
\kappa =  \mmax{\vec{w}}\mmin{\mu}
\{h_{\mu}/ \sqrt{ \vec{w} \cdot \vec{w} } \}~~.
\eeq
The local fields $h_{\mu}$ are given by
\beq
h_{\mu} =  \sigma_\mu \vec{\Psi}\left(\vek{x}^\mu\right) \cdot \vec{w}~~.
\eeq
Here, $\sigma_\mu\in\{-1,1\}$ is the classification of the point
$\vek{x}^\mu$,
for $\mu= 1,\ldots,m$, and $m$ is the total number of labelled examples
in the training set.
This maximization problem is found to be equivalent to a quadratic
minimization
problem for the function $\frac{1}{2} \vec{w} \cdot \vec{w}$ under
the constraints that $h_{\mu} \ge 1 $ for all examples
in the training set.  According to convex optimization theory
the solution vector can be expanded as a linear combination of
example feature vectors via
\beq\label{expans}
\wec =  \sum_\mu \alpha^\mu \sigma_\mu
\vec{\Psi}\left(\vek{x}^\mu\right)
\eeq
where $\alpha^\mu\geq 0$ are Lagrange parameters which account for
the $m$ inequality constraints. Hence, the number of adjustable
parameters $\alpha^\mu$ for this algorithm never exceeds the
number of examples. The $\alpha^\mu$ are nonzero
only for those examples, for which
$ h_{\mu}=  1 $,
defining the {\em support vectors} (SVs) of the data set.
If the remaining examples ($\alpha^\mu= 0$) would be
discarded from the training set, the SVM would predict their
correct label $\sigma_\mu$. Hence, if the relative number of SVs
is small, we can expect that the SVM generalizes well.
In fact, a simple argument \cite{Vap} shows that the
expected ratio of the number of support vectors over $m$ yields an upper
bound on the
generalization error. We will see later within the average case scenario
of Statistical Mechanics that this mechanism
prevents a complex SVM from overfitting when learning a simple rule.

The expansion (\ref{expans}) also reduces the computational cost
of the algorithm
drastically because any inner product of $\wec$ with vectors
$\vec{\Psi}(\vek{x})$ in the feature space
(including $\wec\cdot\wec$) is entirely expressed
in terms of the so called kernel $K(\vek{x},\vek{y})= 
\vec{\Psi}(\vek{x})\cdot \vec{\Psi}(\vek{y})= 
\sum_\rho \Psi_\rho(\vek{x})\Psi_\rho(\vek{y})$.
In particular, for any $\vek{x}$, we have
\beq\label{fields}
\vec{w}\cdot\vec{\Psi}\left(\vek{x}\right)= 
\sum_\mu \alpha^\mu\sigma_\mu
K(\vek{x},\vek{x}^\mu)~~.
\eeq
Hence, both learning and prediction on novel
inputs depend only on the feature vectors $\vec{\Psi}$ through the
kernel $K$.
In fact, there is no need to specify
the high dimensional mapping $\vec{\Psi}(\cdot)$ explicitely.
Instead, one can directly take any reasonable positive semidefinite
operator kernel $K$,
which by Mercer's theorem has a decomposition $K(\vek{x}, \vek{y}) = 
\sum_\rho \lambda_\rho \phi_\rho(\vek{x}) \phi_\rho(\vek{y})$
in terms of eigenvalues $\lambda_\rho$ and orthonormal eigenfunctions
$\phi_\rho(\vek{x})$ and identify $\Psi_\rho$ with
$\sqrt{\lambda_\rho}\phi_\rho$. This approach even allows to take
kernels with feature space dimension $M= \infty$ without problems.

We will now study the generalization performance of SVMs within
the framework of Statistical Mechanics.
We define the partition function
\beq\label{part}
Z =   \int \prod_{\rho=1}^M d w_\rho \, e^{-\frac{\beta}{2}
\vec{w}\cdot\vec{w}}
\prod_{\mu= 1}^m \Theta \!
\left( \sigma_\mu \sum_{\rho=1}^M \sqrt{\lambda_\rho} w_\rho \phi_\rho
(\vek{x}^\mu) - 1 \right)
\eeq
which for $\beta\to\infty$ is dominated by the solution vector $\vec{w}$
of the SVM algorithm.
The properties of the SVM can be computed from the average free energy
$F =  - \frac{1}{\beta} 
{\left\langle\left\langle \ln Z \right\rangle\right\rangle}$,
in the zero temperature limit $\beta\to\infty$,
where the double brackets denote the average over the distribution of
$m$ training examples.
The main difference from the Statistical Mechanics of learning in a
simple perceptron with $M$ weights
is that in the SVM, each coupling $w_\rho$ is weighted by
$\sqrt{\lambda_\rho}$, which typically diminishes the influence of
the more complex, higher order degrees of freedom in the
eigenvector expansion. As we will see, this
makes the generalization behavior of the SVM rather different
from that of a simple perceptron in the thermodynamic limit
$N\to\infty$, when the
rule to be learnt has a similar eigenvector expansion.
We will first consider here a rule of the form
$
\sigma_\mu = 
\sign\left
(\sum_\rho \sqrt{\lambda_\rho} B_\rho \phi_\rho (\vek{x}^\mu) \right)
$
where the teacher weight vector is given by
$B_\rho=\pm 1$.  We will further average the performance over all
teachers of this form
with  equal probability for all nonzero components.
We will specialize on a family of kernels of the form
$K(\vek{x},\vek{y})= k\left(\frac{\vek{x}\cdot\vek{y}}{N}\right)$,
where the only constraint on the function $k(\cdot)$ is the
non-negativity of the eigenvalues.
These kernels are permutation symmetric in the components of the
input vectors and contain the simple perceptron as a special case,
when $k$ is a linear function.
This choice
has the nice feature that for binary input vectors =
$\vek{x}\in\{-1,1\}^N$
the eigenvalue decomposition of $K(\vek{x},\vek{y})$
can be explicitely calculated \cite{Kuehn}.
The eigenfunctions are labelled by subsets
$\rho\subseteq\{1,\ldots,N\}$. We
have $\phi_\rho(\vek{x})= 2^{-N/2}\prod_{i\in\rho} x_i$. The
eigenvalues are
$\lambda_\rho = 
2^{N/2} \sum_{\smvek{x}} K(\vek{e},\vek{x}) \phi_\rho(\vek{x})
$
where $\vek{e}= (1,\ldots,1)^T$, which depend on the cardinality
$|\rho|$ only and show for large $N$ an exponential decay
with $|\rho|$ like
$
\frac{2^N}{N^{|\rho|}} k^{(|\rho|)}(0).
$
The corresponding degeneracy grows exponentially:
$n_{|\rho|}=  {N\choose |\rho|}\simeq N^{|\rho|}/|\rho|!$.

We expect that a decay of the generalization error, $\epsilon_g$, to
zero should occur only on the scale of $m= \ord{M}$,
since $M$ is the number of learnable parameters.
However, as we will show, $\epsilon_g$ may drop to small values already
on a scale
of $m= \alpha N$ examples.
Hence, we make the general ansatz $m= \alpha N^l$, $l\in\N$ and
calculate $f_l =  \lim_{\beta\to\infty}\lim_{N\to\infty} N^{-l} F$.

If we assume that the inputs $\vek{x}^\mu$ are drawn at random with
respect
to a uniform probability distribution $D(\vek{x})$ on $\{-1,1\}^N$, we
can perform the average over the input distribution
by the replica method \cite{Seu91,Waal93,OpKi}.
This becomes tractable by the fact that the eigenfunctions are
orthonormal with respect to $D(\vek{x})$ and we have
$ 2^N \langle \phi_\rho(\vek{x}) \phi_{\rho'} (\vek{x}) \rangle_D
=  \sum_{\smvek{x}} \phi_\rho(\vek{x}) \phi_{\rho'} (\vek{x})
=  \delta_{\rho {\rho'}}$.
Furthermore, all but the constant eigenfunctions have zero mean
under the uniform distribution. By
restricting the kernels to having $k(0)= 0$,
the average over the inputs is expressed in the thermodynamic
limit $N\to\infty$ by expectations over Gaussian random variables.
These averages can be further expressed by the order parameters
\beqas
q_0 & =  & \sum_\rho \Lambda_\rho \<(w_\rho)^2\>~~, \\
q   & =  & \sum_\rho \Lambda_\rho \<w_\rho\>^2~~, \\
R   & =  & \sum_\rho \Lambda_\rho \<w_\rho\> B_\rho
\eeqas
where $\Lambda_\rho = \lambda_\rho/2^N$, and $\<...\>$ denotes a
statistical mechanical averaging specified by Eq.~(\ref{part}).
The generalization error is
$
\epsilon_g =  \frac{1}{\pi} \arccos \frac{R}{\sqrt{Bq}}
$
where $B= \sum_\rho \Lambda_\rho =  k(1) $ is the squared norm of
the teacher vector.
In replica symmetry (which is expected to be exactly fulfilled
by the convexity of the phase space) we obtain
$f_l$ by extremizing the function
\beqa\label{free:two}
f_l(q,R,\chi)= 
\alpha
\int_{-\infty}^{1/\sqrt{q}} \! Dt \, \Phi \!
\left( \frac{Rt}{\sqrt{Bq-R^2}} \right) \frac{(1-\sqrt{q}t)^2}{\chi}
\nonumber \\
+ \frac{1}{2 N^l}
\left( \frac{n_l}{\Lambda^{(+)} -\chi}
+ \frac{1}{\Lambda_l} \right)
\times \\
\times
\left( q - \frac{R^2}{ B^{(-)}
+ n_l\Lambda_l  + \Lambda^{(+)} - \chi }
\right)
\nonumber
\eeqa
with respect to the orderparameters $q, R$ and $\chi$.
Further, $ Dt = \frac{dt}{\sqrt{2\pi}} e^{-t^2/2} $,
$ \Phi(x) = \int_{-\infty}^x Dt $
and $ \chi =  \lim_{\beta\to\infty}\beta(q_0-q) $.
$\Lambda^{(+)} = \sum_{|\rho|>l}\Lambda_\rho$
denotes
the sum over the higher order components
and $B^{(-)} = \sum_{|\rho|<l}\Lambda_\rho$.

As a general result of solving the order parameter equations
we find that all high order components $|\rho|>l$ of the teacher vector
are completely {\em undetermined} by learning only $\ord{N^l}$ examples,
in the sense that
$R^{(+)} =  \sum_{|\rho|>l}\Lambda_\rho w_\rho B_\rho = 0$,
and also that
$q_0^{(+)} =  \sum_{|\rho|>l}\Lambda_\rho (w_\rho)^2 = 0$,
in the large N limit.
However, as we will see, the values of the corresponding weights
$w_\rho$ are not zero
but are determined by the expansion (\ref{expans}).
On the other hand, all lower order components are {\em completely}
determined, in the sense
that $w_\rho= c B_\rho$ for all $|\rho|< l$, where $c$ depends on
$\alpha$ only. The only components which are actually learnt
at a scale $l$ are those for $|\rho|= l$.
We will illustrate these results for quadratic kernels of the
form $k(x)= (1-d)x^2+ dx$, where the parameter $d$, $0<d<1$, tunes
the degree of nonlinearity in the SVM's decision boundary.
On a scale of $m= \alpha N$ examples (left side of Fig.~1),
the SVM is able to learn the linear part of the teacher's rule.
However, since there is not enough information to infer the remaining
$\ord{N^2}$ weights of the teacher's quadratic part, the
generalization error of the SVM reaches a nonzero plateau
with $\epsilon_g(\alpha)-\epsilon_g(\infty) \sim \alpha^{-1}$, where
$\epsilon_g(\infty) = \pi^{-1} \arccos \sqrt{d}$.
This scaling may be understood from the fact that
the undetermined components $w_\rho$ and $B_\rho$, with $|\rho|= 2$
act as a noise term during classification similar to learning of
perceptrons with weight noise \cite{OpKi}.
For comparison, we also show the performance of a simple
{\em linear} SVM (i.e. a perceptron) for which $w_\rho= 0$ when
$|\rho|>1$. The better performance of the nonlinear
SVM does not contradict the fact that, on the linear scale,
its higher order weights
$w_\rho$ for $|\rho|= 2$ are uncorrelated with the corresponding
teacher values. Those weights are needed
to learn the training examples perfectly which is not possible
for the linear machine when $\alpha$ exceeds a critical value
$\alpha_c(d)$,
given by $ \pi/\alpha_c =  \arctan \pi/(\alpha_c d) $.

\begin{figure}[ht]
\begin{center}
\setlength{\unitlength}{1mm}
\begin{picture}(85,60)
\put(0,0){\makebox(85,60)
          {\includegraphics{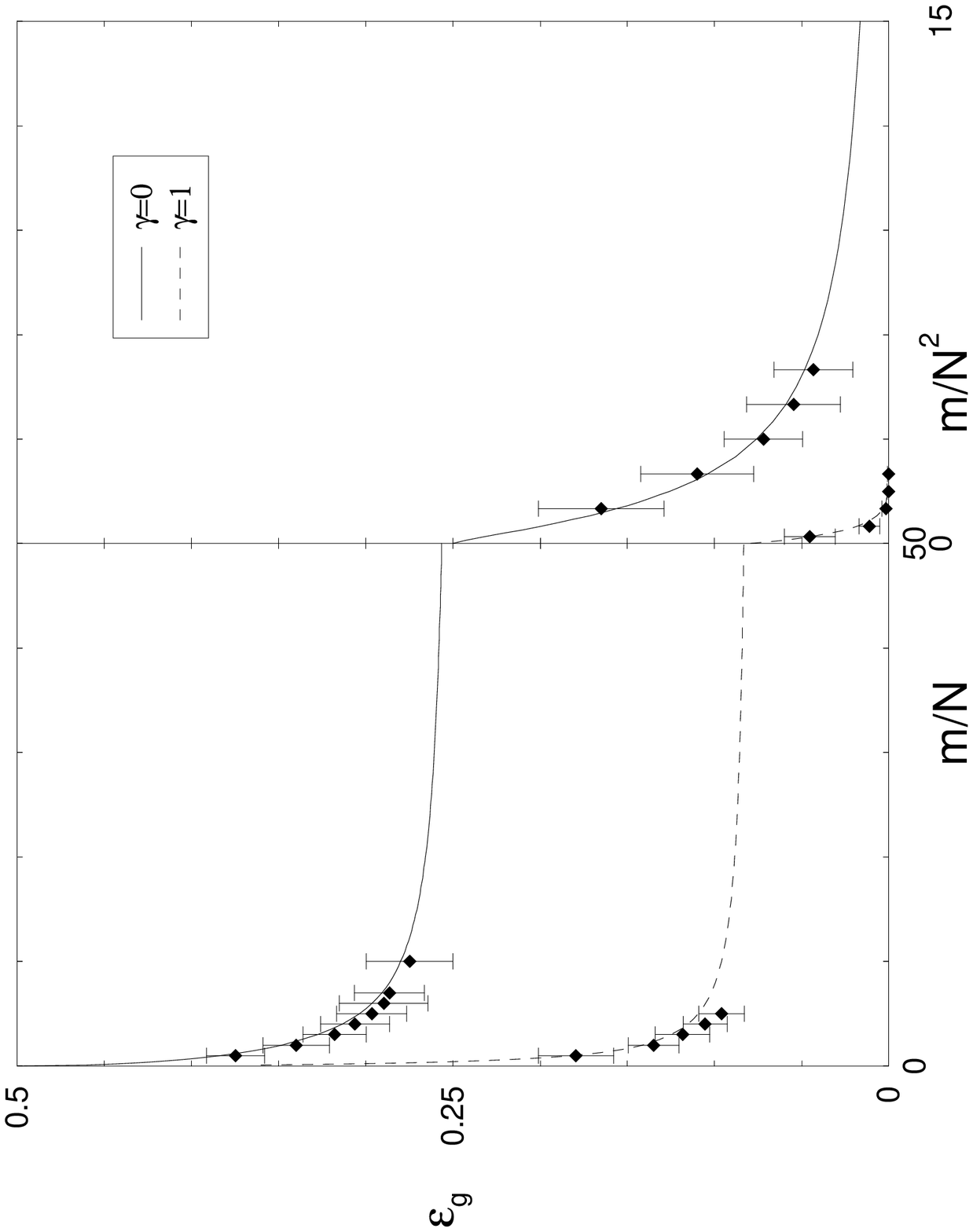}}}
\put(1,0){\makebox(85,60)
          {\includegraphics{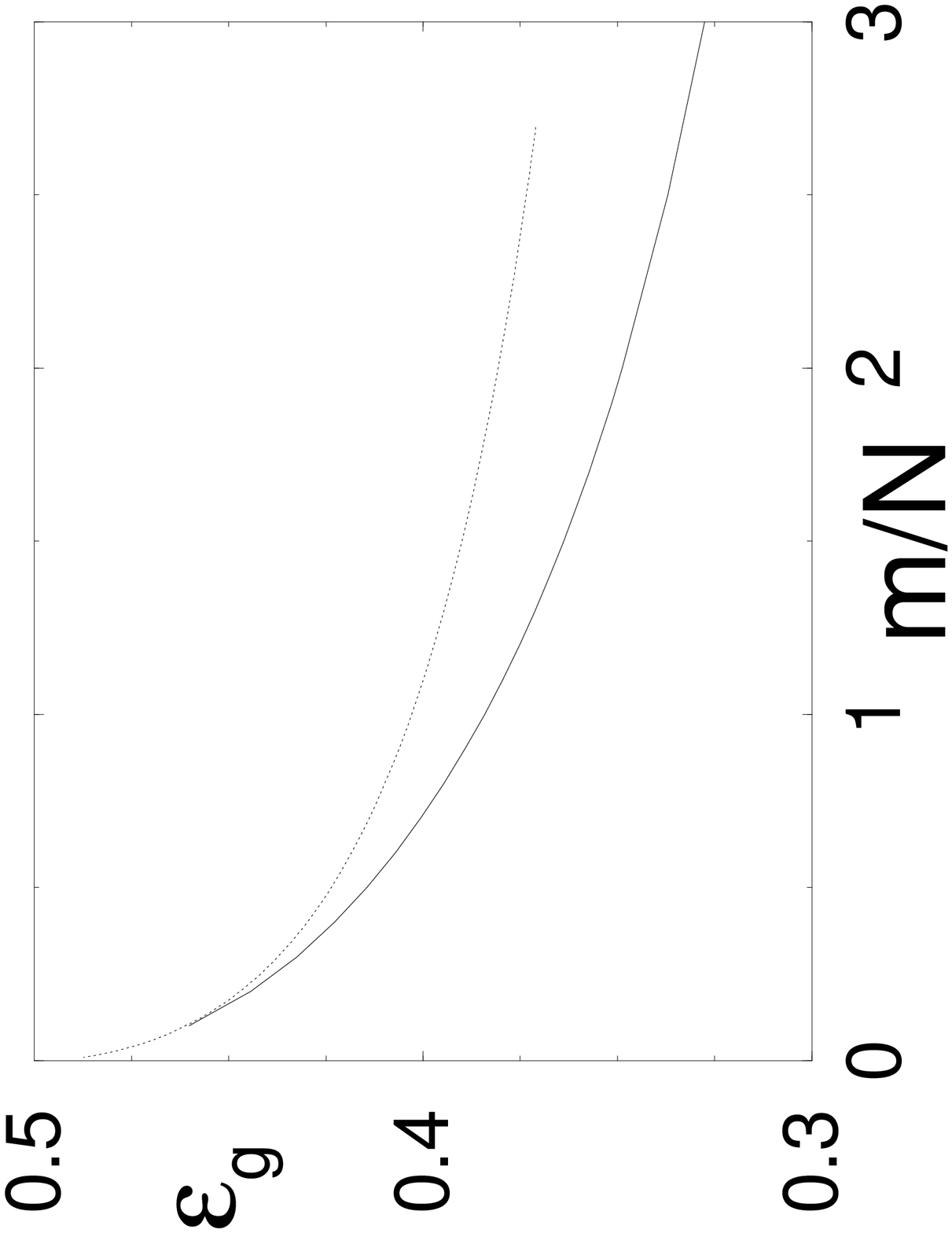}}}
\end{picture}
\end{center}
\caption{
Decrease of the generalization error on different scales of examples,
for quadratic SVM-kernel learning a quadratic teacher rule ($d= 0.5, \, B=1$)
and various gaps $\gamma$.
The inset compares the SVM to a linear perceptron (upper curve) trying to
learn the same task.
Simulations were performed with N= 201 and averaged over 50 runs (left and
next figure),
and N= 20, 40 runs (right).
}
\end{figure}

Increasing the number of examples to a scale
of $m= \alpha N^2$ (right side of Fig.~1), the well known
\cite{OpHa95} $1/\alpha$ asymptotic vanishing of $\epsilon_g$ is found.
A similar stepwise learning has been obtained for the case of
Gibbs learning in higher order
perceptrons \cite{Yoon98}. In general, for kernels which are polynomials of
order $z$, more plateaus will appear. On the scale of $m= \alpha N^{l-1}$
examples, the generalization error decays to a plateau
at $\alpha\to\infty$ given by
\beq
\epsilon_g = 
\frac{1}{\pi} \arccos
\sqrt{\frac{B^{(-)}}{B}}= 
\frac{1}{\pi} \arccos \sqrt{\frac{ \sum_{j= 1}^{l-1}
\frac{k^{(j)}(0)}{j!} }
{k(1)}}.
\eeq
Finally, at the highest scale $m= \alpha N^z$, the generalization error
converges to zero as
$\epsilon_g \approx \frac{0.500489}{z!}\alpha^{-1}$.
This form is in accordance with general results \cite{Vap} which show that
(in the worst case) the number of examples must be larger
than the capacity of the classifier in order to achieve a small
generalization error.
The capacity $m_c= \alpha_c N^z$ is found from (\ref{free:two})
by solving the order parameter equations with the restriction $R= 0$,
as the value of $\alpha$ where $q_0$ diverges. We obtain
$\alpha_c= \frac{2}{z!}$ which agrees with the results in \cite{cover}
for polynomial separation surfaces in the large $N$ limit.

As the next problem, we study the ability of the SVM
to cope with the problem of overfitting when learning a simple rule.
We keep the SVM quadratic, but choose a simpler,
linear teacher rule according to
$|B_\rho|= 1$ for $|\rho|= 1$ and $|B_\rho|= 0$ else.
The results for the generalization error,
obtained by a straightforward extension of (\ref{free:two}),
are shown in Fig.~2,
where the number of examples is scaled as $m= \alpha N$.
Surprisingly, although the student has of $\ord{N^2}$ adjustable parameters,
this does not lead to any strong overfitting.
The SVM is able to learn the $N$ teacher weights on the scale
of $m= \alpha N$ examples far below capacity. For comparison, we have also
shown $\epsilon_g$ for a simple linear SVM (i.e. with $w_\rho= 0$
for $|\rho|= 2$). While
for the latter case, the decay of the generalization error
is of the well known form $\epsilon_g\sim \alpha^{-1}$, the
quadratic SVM shows
the somewhat slower decay $\epsilon_g\sim \alpha^{-2/3}$.
The same scaling is obtained for higher order SVMs which learn a low order
{\it e.g.,} a linear,  rule.

\begin{figure}[ht]
\begin{center}
\setlength{\unitlength}{1mm}
\begin{picture}(85,60)
\put(0,0){\makebox(85,60)
          {\includegraphics{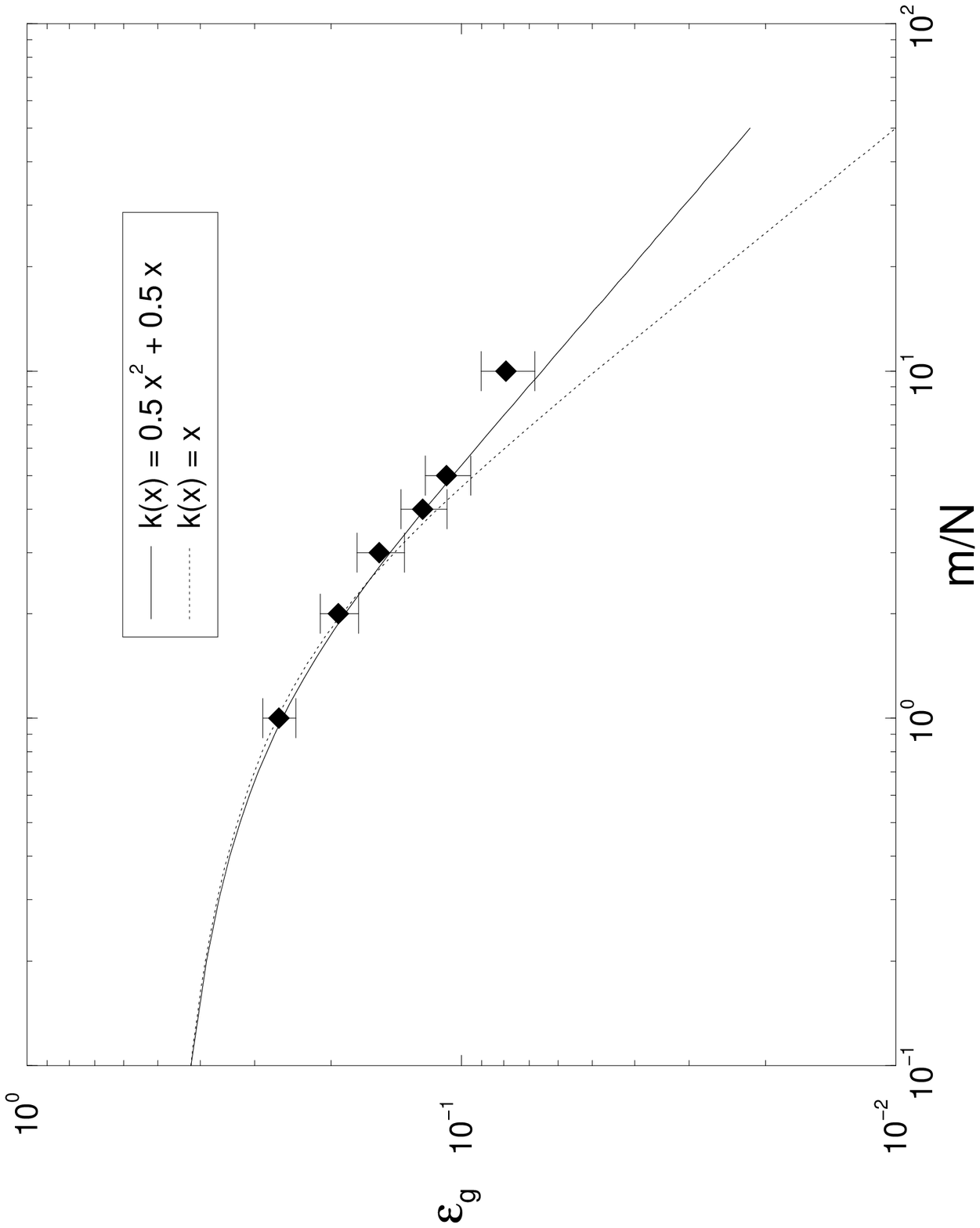}}}
\put(1,0){\makebox(85,60)
          {\includegraphics{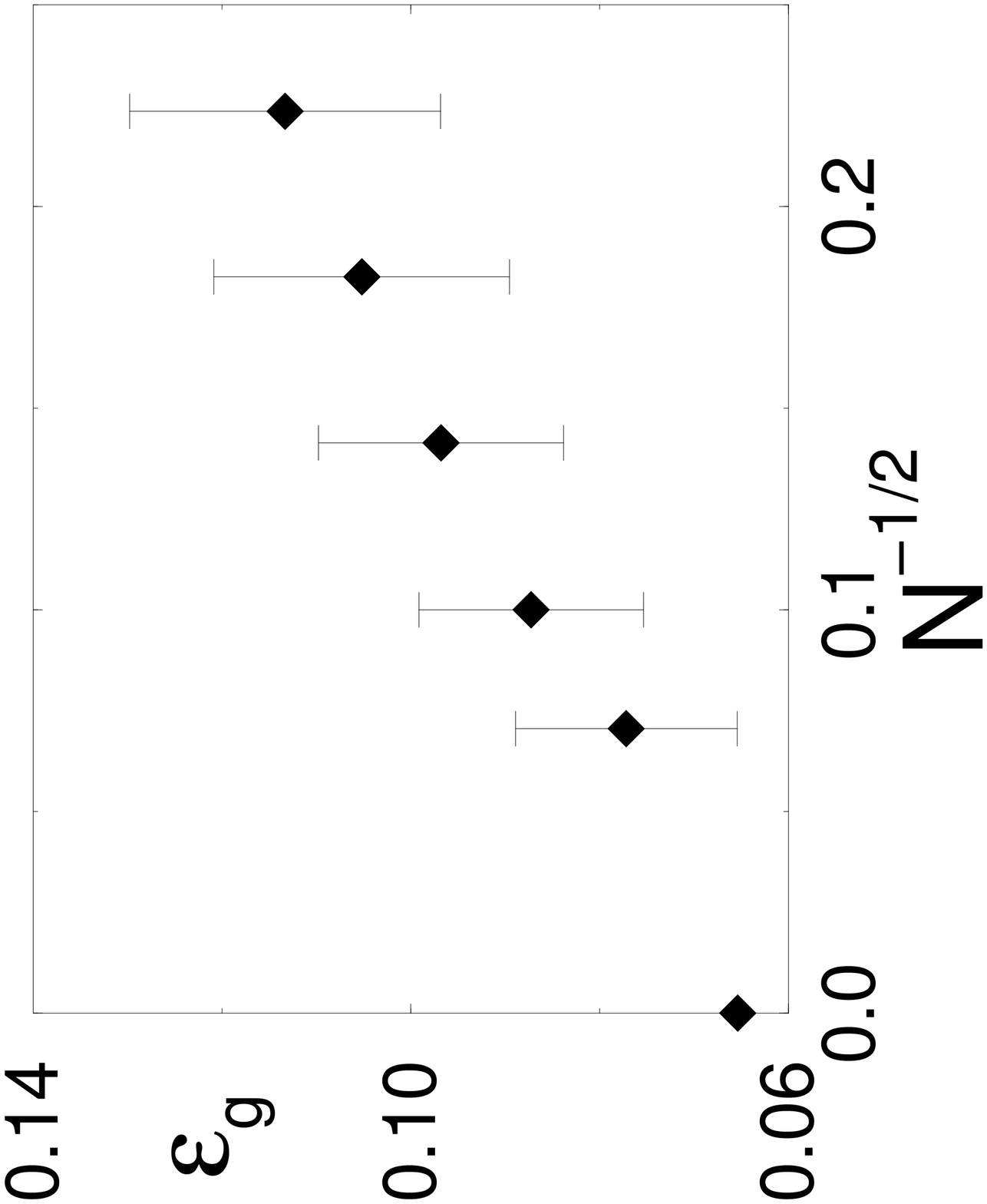}}}
\end{picture}
\end{center}
\caption{
Learning curves for linear student and quadratic SVM-kernels, all learning
a linear teacher rule ($B= d$). For $\alpha= 10$, a finite size scaling is
shown as inset.
}
\end{figure}

We can shed further light on this interesting result by
showing that the number of SVs
increases like $\alpha^{2/3}$, hence the relative number of SVs
(which is a crude upper bound on $\epsilon_g$)
{\em decreases} like $\alpha^{-1/3}$. This can be
understood from the following analysis, which is valid for more general 
classes of input distributions. For simplicity, we restrict ourselves to the
quadratic SVM learning a linear rule. We assume that the inputs have zero mean 
and are sufficiently weakly correlated such that the offdiagonal elements of 
the quadratic part of the kernel matrix
$K_{\mu\nu}^{(2)} =  (1-d) (N^{-1}\vek{x}^\mu \cdot \vek{x}^\nu)^2$
for $\mu\neq\nu$ are typically $\ord{1/N}$. The diagonal elements are
$K_{\mu\mu}^{(2)} =  1-d $.  Evaluating
$h_\mu= \sigma_\mu \vec{\Psi}\left(\vek{x}^\mu\right) \cdot \vec{w}$ 
using Eq.~(\ref{fields}) one finds that
the relative contributions of the off-diagonal elements of $K$
are $\ord{(m/N^2)^{\oneh}}$ and can be neglected on the linear scale
$m= \alpha N$.  Hence we obtain
$h_\mu= v_{\mu}+ (1-d)\alpha^\mu$
with $v_{\mu}$ being the contribution from the
linear weights, namely, 
$v_{\mu} =  \sigma_{\mu} \sqrt{d/N} \vek{w} \cdot \vek{x}$, 
where $\vek{w}$ consists only of $w_\rho$ with $|\rho|=1$. Solving for
the coefficients $\alpha^\mu$, noting that they are nonzero only when
$h_\mu =1$,  we obtain
\beq\label{alpha}
\alpha^\mu =  (1-d)^{-1}(1- v_\mu)\Theta(1- v_\mu)~~.
\eeq
When $\alpha$ is small, all $\alpha^\mu \approx 1/(1-d)$
and the SVM acts like a Hebbian classifier.
With increasing number of examples $v_\mu$ will grow and
the probability that $\alpha^\mu>0$ (an example is a SV)
will decrease.  The exact asymptotic scaling can be
calculated selfconsistently assuming that for large $\alpha$,
$w_\rho\simeq c B_\rho$ for $\rho= 1$ and
$c= N^{-1} \sum_{|\rho|= 1} w_\rho B_\rho = 
\frac{1}{N}\sum_{\mu= 1}^{\alpha N}\alpha^\mu u_\mu$
where
$u_\mu $ is the linear contribution to the local field of the teacher vector.
Using Eq.~(\ref{alpha}) and noting that $v_\mu \approx c u_\mu$ we obtain
\beq\label{c}
c \sim \alpha \int_0^{1/c} du~ p(u)\,u\,(1-cu)
\eeq
valid for large $\alpha$.
Here $p(u)$ denotes the density of the teacher linear fields $u$.
Solving Eq.~(\ref{c}) for $c$ in limit of $\alpha\to\infty$ yields 
$c \sim (\alpha p(0)/6)^{1/3}$.
Similarly, the relative number of SVs scales as
$p(0)/c\sim\alpha^{-1/3} p(0)^{2/3}$.

The dependence on $p(0)$ suggests that the density of inputs
at the teacher's decision boundary
should play a crucial role for the generalization ability of the SVM.
When this density vanishes close to
the teacher's separating hypersurface, a much faster
decay of the generalization error can be expected.
To study this property in more detail, we have analyzed the Statistical
Mechanics
for an input distribution correlated with the teacher weights such that
$D(\vek{x}) \sim
\Theta\left( \sigma \sum_\rho \sqrt{\Lambda_\rho} B_\rho \phi_\rho
(\vek{x}) - \gamma \right)$ 
which has a gap of zero density with size $2\gamma$ around the teacher's 
decision boundary.
As expected, the generalization performance of a quadratic SVM which learns 
from a quadratic teacher is enhanced, but the asymptotic decay towards the 
plateau on the linear scale (see Fig.~1) is still of the form
$\epsilon_g(\alpha)-\epsilon_g(\infty)\sim\alpha^{-1}$.
The effect of the gap is more dramatic on the highest scale $m = \alpha N^2$, 
where instead of an inverse power law, we now find a fast drop of the 
generalization error like
$\epsilon_g \sim \alpha^{-3} e^{-\hat{c}(\gamma)\alpha^2}$.
\\ \\

{\bf Acknowledgements:} We would like to thank H.~Yoon and J.H.~Oh
for sending us their paper \cite{Yoon98} before publication.
The work was supported by
the grant (Op 45/5-2) of the Deutsche Forschungsgemeinschaft.
The work of HS was supported in part by the USA-Israel Binational 
Science Foundation.


\begin{references}

\bibitem{Seu91} H. Seung, H. Sompolinsky, and N. Tishby,
	Physical Review A {\bf 45}, 6056 (1992).
\bibitem{Waal93} T. L. H. Watkin, A. Rau and M. Biehl,
	Rev. Mod. Phys. {\bf 65}, 499 (1993).
\bibitem{OpKi} M. Opper and W. Kinzel,
	{\em Statistical Mechanics of Generalization},
	in {\em Physics of Neural Networks III},
	ed. by J. L. van Hemmen, E. Domany and K. Schulten,
	(Springer Verlag, Berlin, 1996).
\bibitem{Monass} R. Monasson and R. Zecchina, Phys. Rev. Lett. {\bf 75}, 
	2432 (1995).
\bibitem{Vap} V. N. Vapnik, {\em The Nature of Statistical
	Learning Theory}, Springer Verlag, New York (1995).
\bibitem{Cort} C. Cortes and V. Vapnik, Machine Learning {\bf 20}, 
	273-297 (1995).
\bibitem{Kuehn} R. K\"uhn and J. L. van Hemmen,
	in {\em Physics of Neural Networks I},
	ed. by J. L. van Hemmen, E. Domany and K. Schulten,
	(Springer Verlag, Berlin, 1996).
\bibitem{OpHa95} M. Opper and D. Haussler, Phys. Rev. Lett. {\bf 75}, 
	3772 (1995);
	Europhys. Lett. 37, 31 (1997).
\bibitem{Yoon98} H. Yoon and J. H. Oh, J. Phys. A {\bf 31}, 7771 (1998). 
\bibitem{cover} T. Cover; IEEE Trans. El. Comp. 14, 326 (1965).

\end{references}
\end{document}